\documentclass[doublecol]{epl2} 

\usepackage{amssymb,amsmath,dsfont}

\title{Amplitude bounds for biochemical oscillators}

\author{David J. J\"{o}rg\thanks{E-mail: \email{djj35@cam.ac.uk}}}
\shortauthor{David J. J\"{o}rg}

\institute{                    
Theory of Condensed Matter Group, Cavendish Laboratory, University of Cambridge \\ -- JJ Thomson Avenue, Cambridge CB3 0HE, UK\\
The Wellcome Trust/Cancer Research UK Gurdon Institute, University of Cambridge \\ -- Tennis Court Road, Cambridge CB2 1QN, UK
}

\pacs{05.45.-a}{Nonlinear dynamics}
\pacs{87.10.Ed}{Differential equations in mathematical aspects of biological physics}
\pacs{82.39.-k}{Chemical kinetics in biological systems}

\abstract{We present a practical method to obtain bounds for the oscillation minima and maxima of large classes of biochemical oscillator models that generate oscillations through a negative feedback. These bounds depend on the feedback nonlinearity and are independent of explicit or effective feedback delays. For specific systems, we obtain explicit analytical expressions for the bounds and demonstrate their effectiveness in comparison with numerical simulations.}

\begin{document}

\maketitle

\section{Introduction}
Oscillations can arise when a dynamical system forms a negative-feedback loop with itself~\cite{strogatz01}.
Through periodic changes, the paradox of self-negation is resolved if the negative feedback is accompanied by a sufficiently large time delay~\cite{kauffman2002}.
Such negative-feedback oscillations are ubiquitous in living systems---a prominent example are biochemical oscillations, i.e., periodic changes in the concentration of gene products  within cells~\cite{goldbeter97,morelli07,ferrell13,novak08}.
The negative feedback of a gene onto itself is typically mediated by intermediary components (such as mRNA and proteins) with finite synthesis times and lifetimes, which provide the delays necessary for oscillations to occur~\cite{novak08}.
An additional way to create feedback delays is through indirect self-interactions that arise as a consequence of coupling between oscillatory elements~\cite{jensen09,chakraborty12}.
Biochemical oscillations are used by living organisms as biological pacemakers and clocks that govern vital functions during development and life, e.g., daily rhythms~\cite{bargiello84,zehring84,dunlap99,smolen02,schibler05,hastings07,zwicker10}, DNA~replication~\cite{novak97}, the cell cycle~\cite{gerard11}, neuronal differentiation~\cite{imayoshi14,shimojo16b}, and embryonic pattern formation~\cite{lewis03,oates12,schroter12,jorg15}.
Moreover, in recent years, different synthetic biochemical oscillators have been engineered~\cite{elowitz00,garciaojalvo04,stricker08,danino10,kim11,niederholtmeyer15,potvintrottier16}, which has lead to an increased interest in theoretical approaches to characterise such oscillations~\cite{elowitz00,hasty02,garciaojalvo04,ullner08}. 

Apart from their period, an important functional feature of such oscillations is their amplitude.
Since theoretical models of negative-feedback oscillators must comprise a nonlinearity and a delay in the feedback---either explicitly or through intermediate products~\cite{novak08}---analytical solutions are often not possible.
Analytical amplitude estimates for specific systems typically require a separate treatment of different parameter regimes, e.g., regions close to the Hopf bifurcation~\cite{woller14} or the limit of strong feedback~\cite{woller13}.
Hence, it is often elusive how  the details of the nonlinear feedback govern the oscillation amplitude.

In this paper, we present a rigorous but practical method to derive bounds for the minima and maxima of periodic solutions for large classes of biochemical oscillator systems, including systems with explicit or effective delays as well as spatially extended systems.
Using specific systems as examples, we illustrate our results and demonstrate the effectivity of the derived bounds in comparison with numerical simulations.

\section{Biochemical oscillators with delayed feedback}
\begin{figure}
\onefigure[width=87mm]{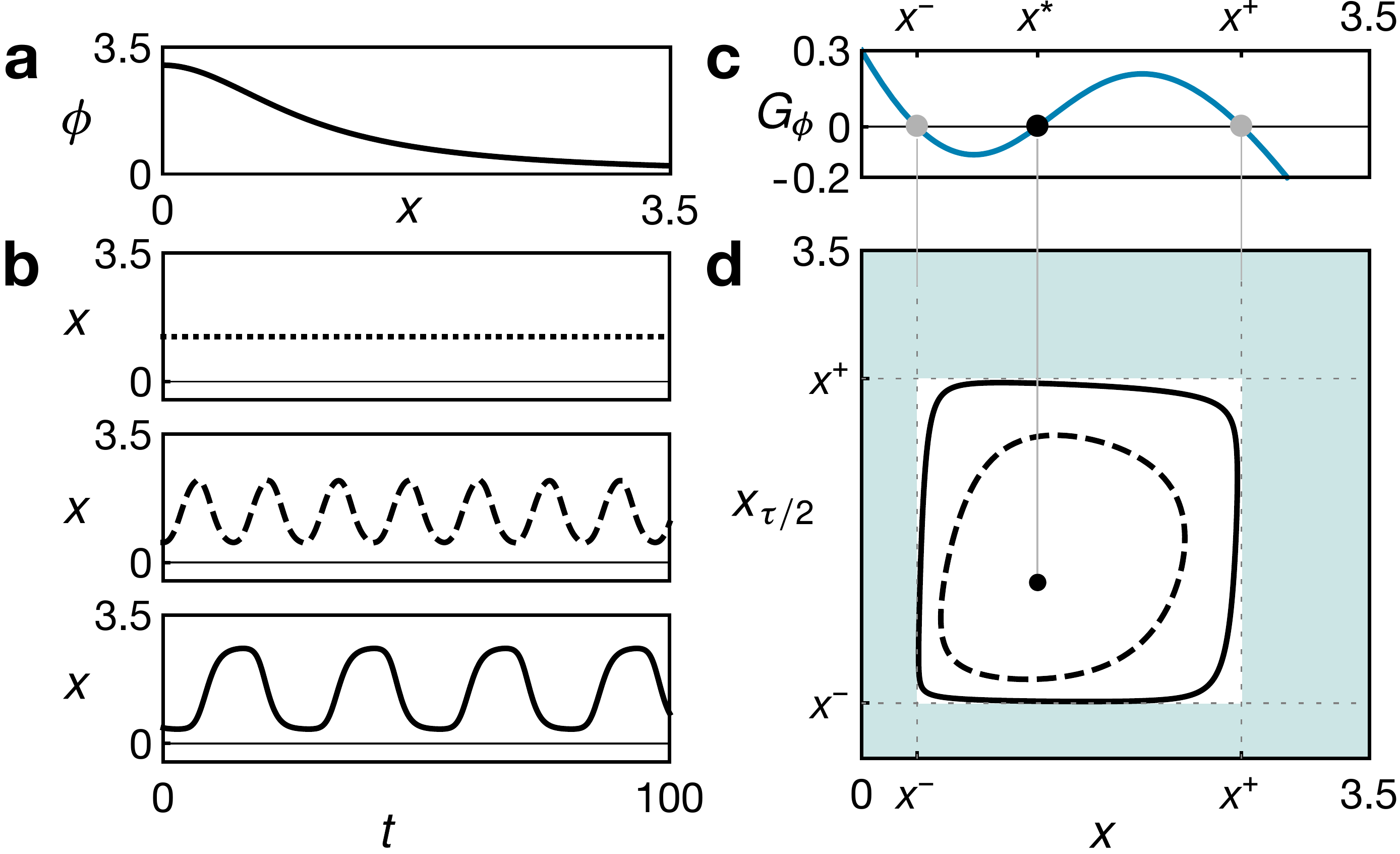}
\caption{Examples for fixed points and limit cycles of eq.~(\ref{eq.dyn}). (a) Hill-type feedback function $\phi(x)=3/(1+x^2)$ used for the examples shown in panels b and d. (b) Trajectories for $x(t)$ for systems with different feedback delays: $\tau=1$ (dotted), $\tau=6$ (dashed), $\tau=12$ (solid).
(c) The function $G_\phi$ as defined in eq.~(\ref{eq.psi}) with $\phi$ as specified for panel a. Dots indicate the roots of $G_\phi$.
(d) Parametric representation of the same trajectories as in panel b with the variable $x$ plotted versus the delayed variable $x_{\tau/2}$, where $\tau$ is the feedback delay. Dotted lines show the values $x^+$ and $x^-$, i.e., the largest and smallest root of $G_\phi$.}
\label{fig.cycles}
\end{figure}

As a starting point, we consider a widely used model for a biochemical oscillator: a single variable $x$ (representing the concentration of a biochemical component or a gene activity) that satisfies the delay-differential equation~\cite{mackeyglass77,novak08}
\begin{equation}
	\dot x = \phi(x_\tau)-x  \ , \label{eq.dyn}
\end{equation}
where $x_\tau(t)=x(t-\tau)$ with $\tau$ being the feedback delay. 
The biochemical feedback of $x$ onto itself is described by a monotonically decreasing feedback function $\phi$ and the linear term describes the decay of $x$.
Here, we consider the non-dimensional form of the equation with unit decay rate.
Due to the presence of the delay $\tau$, eq.~(\ref{eq.dyn}) constitutes an infinite-dimensional dynamical system~\cite{hale93}.
For appropriate choices of $\phi$, such systems can exhibit stable limit cycle oscillations if the delay $\tau$ exceeds a critical value that marks a Hopf bifurcation~\cite{wei07,novak08}, see figs.~\ref{fig.cycles}a,b for examples.
Numerically, it is found that for typical choices of feedback functions $\phi$, the oscillation amplitude saturates as the feedback delay is increased to values $\tau \gg 1$, see solid curve in fig.~\ref{fig.cycles}b,d.

Which features of the system impose these amplitude bounds and under which circumstances can we find explicit expressions for the bounds?

\section{Heuristics}%
Candidates for amplitude bounds can be found using a simple self-consistency argument. Any solution $\tilde x(t)$ to eq.~(\ref{eq.dyn}) trivially satisfies the one-dimensional non-autonomous equation $\dot {\tilde{x}}(t) = \varphi(t) - \tilde x(t)$ with $\varphi(t) \equiv \phi(\tilde x_\tau(t))$. 
If $\varphi$ is approximately constant over a given period of time, then $\tilde x$ converges exponentially towards the corresponding value of $\varphi$.
Indeed, for the case of large delays, we find that $\tilde x$ spends extended periods of time near the oscillation minimum $x^-$ and maximum $x^+$ (see, e.g., solid curve in figs.~\ref{fig.cycles}b,d); consequently, $\varphi$ inherits this nearly constant behaviour a time $\tau$ later.
Since the feedback is negative, the system exponentially converges towards $x^+$ when the feedback responds to $x^-$ in the past and vice versa. This suggests that in such cases, $x^+$ and $x^-$ are related by
\begin{equation}
	\phi(x^\pm) = x^\mp \ . \label{eq.def.bounds}
\end{equation}
In the case that such approximately constant periods do not exist (see, e.g., the dashed curve in figs.~\ref{fig.cycles}b,d), the system still tries to continuously follow the trajectory $\varphi(t)$ but is unable to converge. This is the case when the rate of change of $\varphi$ is comparable to the exponential convergence rate, which is set by the decay rate. This self-consistently constrains the periodic solution.
Importantly, Eq.~(\ref{eq.def.bounds}) implies that $x^+$ and $x^-$ can be found among the real, non-negative roots of the function
\begin{equation}
	G_\phi(x) = \phi^2(x)-x \ , \label{eq.psi}
\end{equation}
where $\phi^2(x)=\phi(\phi(x))$, see fig.~\ref{fig.cycles}c,
which entails a practical way to analytically or numerically determine these values, as shown later on.
Note that the preceding considerations rely entirely on self-consistency arguments and use already known properties of the periodic solution to claim that eq.~(\ref{eq.def.bounds}) defines bounds for the oscillation minima and maxima.
We now formulate three propositions to make this claim precise.

\section{Proposition~I}%
Let $\phi(x)$ be a smooth function with $\phi \geq 0$ and $\phi'|_{x\geq 0} \leq 0$, such that $G_\phi$ as defined in eq.~(\ref{eq.psi}) has a finite number of real, positive roots with the smallest root $x^-$ and the largest root $x^+$ satisfying $x^+ \neq x^-$. Then, $\phi(x^\pm)=x^\mp$.

\emph{Proof.} Smoothness, monotonicity, and non-negativity imply that $\phi$ has exactly one real, non-negative invariant point $x^* = \phi(x^*)$. Obviously, $x^*$ is a root of $G_\phi$. Hence, for any real, positive root $y \neq x^*$ of $G_\phi$, uniqueness of $x^*$ implies that $y' \equiv \phi(y)$ satisfies $y' \neq y$ and $\phi(y')=y$, implying that $y'$ is another root of $G_\phi$ with $y' \neq x^*$. Hence, $G_\phi$ must have an odd number of roots with all roots apart from $x^*$ belonging to a unique pair such that each pair $(y,y')$ is related by $\phi(y)=y'$ and $\phi(y')=y$. Therefore, the existence of two roots $x^+ \neq x^-$ implies that $G_\phi$ must have at least three roots and since $x^+$ and $x^-$ are the largest and smallest root, respectively, it follows by monotonicity of $\phi$ that they form such a pair.

\section{Proposition~II}%
For any periodic solution $\tilde x(t)$ of eq.~(\ref{eq.dyn}) with a bounded feedback function $\phi$ satisfying the requirements of Proposition~I, let $\mathcal{C}=\{ \tilde x(t) \ | \ t \}$ be the set of all values of $\tilde x$. Then, $\mathcal{C} \subseteq \phi(\mathcal{C})$ where $\phi(\mathcal{C}) = \{ \phi(x) \ | \ x \in \mathcal{C} \}$ is the image set of $\mathcal{C}$ under $\phi$.

\textit{Proof.}
We transform  eq.~(\ref{eq.dyn}) into an integral equation by evaluating $\smash{\int_0^t \mathrm{e}^{-(t-s)} \dot x(s) \, \upd s}$ and partially integrating,
\begin{equation}
	 x(t) = x(0) \mathrm{e}^{-t} + \smash{\int_0^t \mathrm{e}^{-(t-s)} \phi( x_\tau(s)) \, \upd s}  \ .
\end{equation}
For the periodic solution $x(t)=\tilde x(t)$, it follows from $\phi(\tilde x_\tau) \in \phi(\mathcal{C})$ and the boundedness of $\phi$ that
\begin{equation}
	\tilde x(t) \leq \tilde x(0) \mathrm{e}^{-t} + (1-\mathrm{e}^{-t}) \max \phi(\mathcal{C}) \ . \label{ineq}
\end{equation}
The long-time behaviour of the inequality~(\ref{ineq}) and the periodicity of $\tilde x$ imply that all $x \in \mathcal{C}$ satisfy the bound $x \leq \max \phi(\mathcal{C})$.
The proof for $x \geq \min \phi(\mathcal{C})$ for all $x \in \mathcal{C}$ follows analogously. This shows $\mathcal{C} \subseteq \phi(\mathcal{C})$.

\section{Proposition~III}%
Let $\phi$ satisfy the requirements of Propositions~I and II and let $\mathcal{C}^* = [x^-,x^+]$. Then, the implication $\mathcal{L} \subseteq \phi(\mathcal{L}) \Rightarrow \mathcal{L} \subseteq \mathcal{C}^*$ holds for any real interval $\mathcal{L}=[\ell^-,\ell^+]$.

\emph{Proof.} We first show the inequalities

(a) $\phi^2|_{x>x^+} < x$,

(b) $\phi^2|_{0 \leq x<x^-} > x$.

Inequality (a) follows from the smoothness of $\phi$ and the fact that $x^+$ is the largest root of $G_\phi$; therefore $\operatorname{sign}(G_\phi)|_{x>x^+}=\mathrm{const.}$ and the alternative $\phi^2|_{x>x^+} > x$ would contradict the boundedness of $\phi$.
Inequality (b) follows from the smoothness of $\phi$, the fact that $x^- > 0$ is the smallest root of $G_\phi$ and $G_\phi(0)=\phi^2(0) \geq 0$.

\begin{figure}
\onefigure[width=87mm]{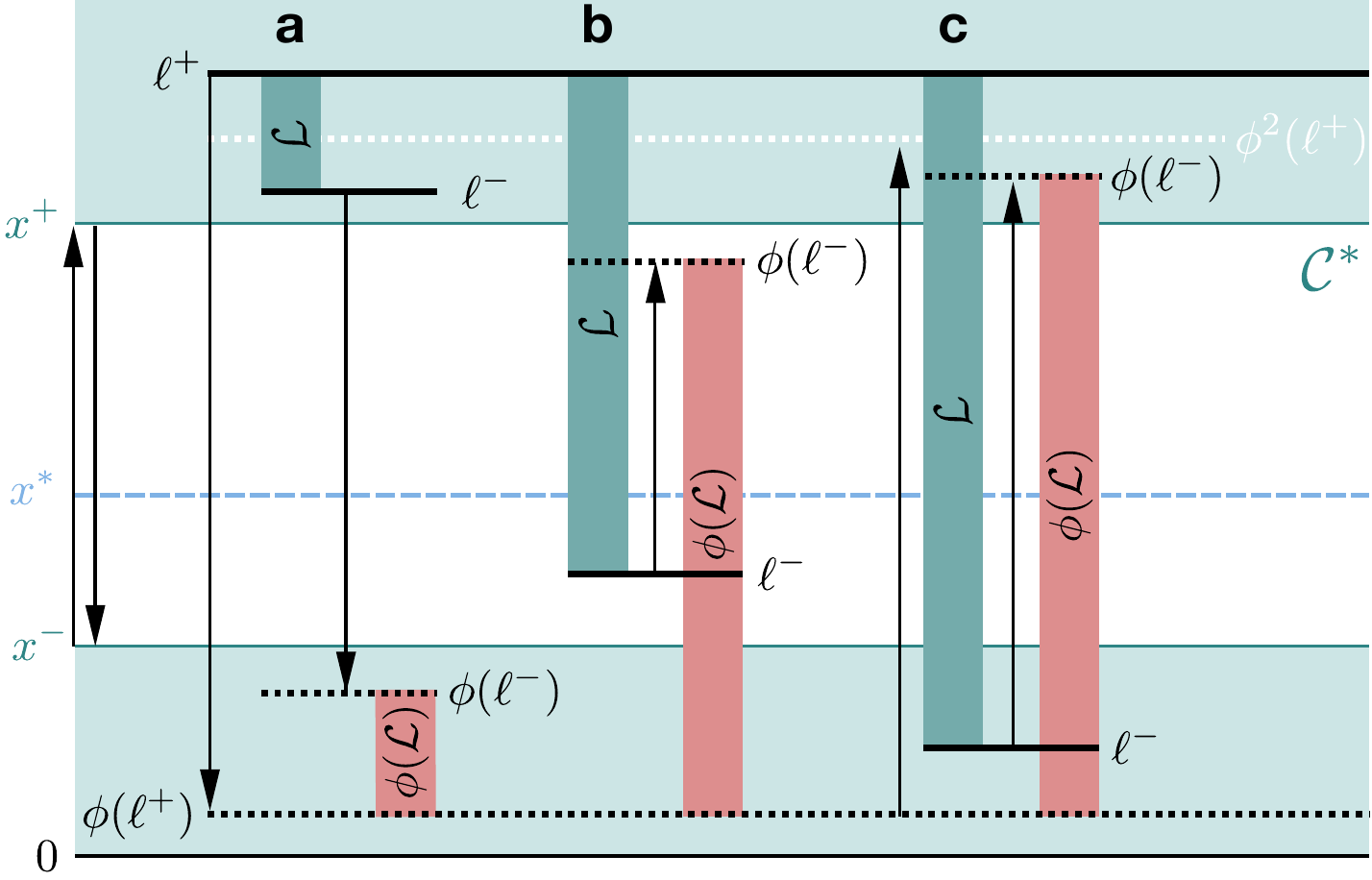}
\caption{Graphical representation of the contradictions that follow from the assumption $\mathcal{L} \subseteq \phi(\mathcal{L})$ but $\mathcal{L} \not\subset \mathcal{C}^*$. Arrows indicate the action of $\phi$ on the bounds of an interval $\mathcal{L}=[\ell^-,\ell^+]$ with $\ell^+ > x^+$ for the cases (a) $\ell^- > x^+$, (b) $\ell^- \in \mathcal{C}^*$, (c) $\ell^- < x^-$.}
\label{fig.proof}
\end{figure}

The proof proceeds by contradiction and we therefore assume $\mathcal{L} \subseteq \phi(\mathcal{L})$ but $\mathcal{L} \not\subset \mathcal{C}^*$.
The latter assumption implies that at least one of the bounds $\ell^+ \leq x^+$ and $\ell^- \geq x^-$ is violated.
Without loss of generality, consider the case $\ell^+ > x^+$, see fig.~\ref{fig.proof}. Since $\ell^- \geq \phi(\ell^+)$ and $\ell^- \geq 0$ follow from $\mathcal{L} \subseteq \phi(\mathcal{L})$, we obtain $\phi(\ell^-) \leq \phi^2(\ell^+) < \ell^+$ via the monotonicity of $\phi|_{x\geq 0}$ and inequality (a), implying the contradiction $\mathcal{L} \not\subset \phi(\mathcal{L})$.
The case $\ell^- < x^-$ follows analogously using inequality (b).

\section{Remarks}
The three propositions show how, under minimal monotonicity and regularity requirements for $\phi$, the largest and smallest real, positive roots $x^+$ and $x^-$ of $G_\phi$, eq.~(\ref{eq.psi}), bound the amplitude of any periodic solution $\tilde x$ of eq.~(\ref{eq.dyn}): Proposition~I implies that $\phi$ alternates between the two bounds, a property expressed by eq.~(\ref{eq.def.bounds})\footnote{Note that eq.~(\ref{eq.def.bounds}) can be interpreted as describing period-2 oscillations of the time-discrete system $x_{i+1}=\phi(x_i)$. It has been shown previously that the existence of such discrete oscillations and the existence of continuous oscillations of eq.~(\ref{eq.dyn}) are linked~\cite{enciso07}.}, and Proposition~II and III show the boundedness of periodic solutions via $\mathcal{C} \subseteq \phi(\mathcal{C})$ and the implication $\mathcal{C} \subseteq \phi(\mathcal{C}) \Rightarrow \mathcal{C} \subseteq \mathcal{C}^*$, where $\mathcal{C}$ is the set of values of the periodic solution and $\mathcal{C}^*=[x^-,x^+]$.

\begin{figure*}
\onefigure[width=180mm]{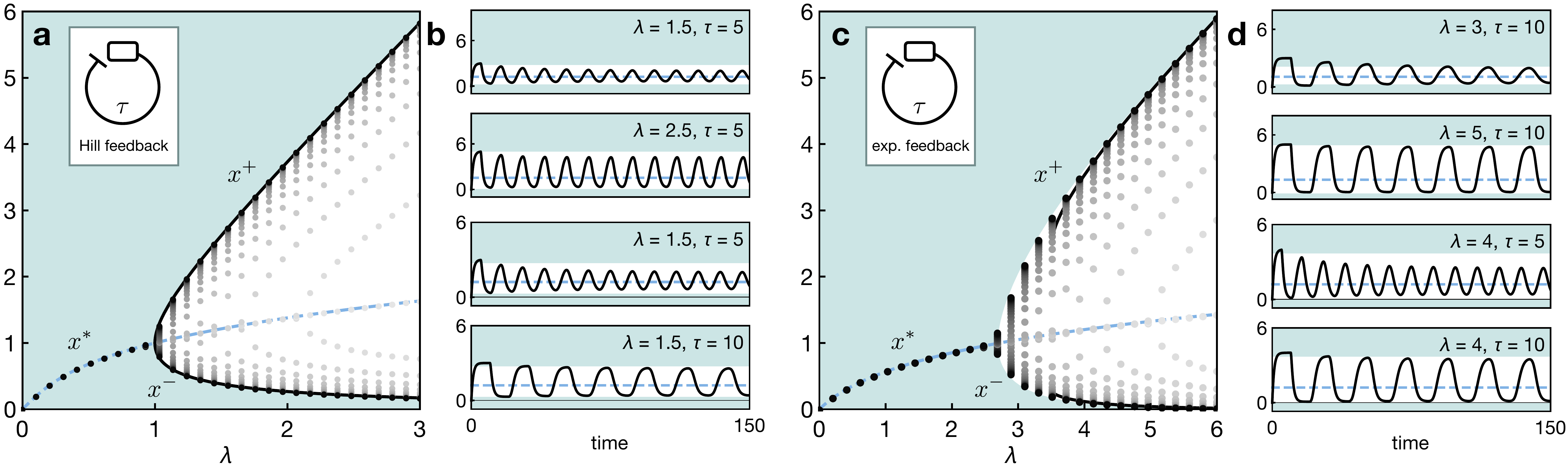}
\caption{Analytical bounds and numerical solutions for the delay system  eq.~(\ref{eq.dyn}).
(a,b)~Systems with Hill-type feedback, given by eq.~(\ref{eq.hill}) with $n=2$.
(c,d)~Systems with exponential feedback, given by eq.~(\ref{eq.exp}).
(a)~Bounds $x^+$ and $x^-$ (solid black curves) and the invariant point $x^*$ (dashed blue curve) as a function of $\lambda$, as given by eqs.~(\ref{eq.hill.amp},\ref{eq.hill.eq}). Dots show minima and maxima of the long-time periodic dynamics of numerical solutions of eq.~(\ref{eq.dyn}) with different delays ($\tau=0,\hdots,20$ from bright to dark).
(c) Bounds and invariant points analogous to panel a. Solid black curve: eq.~(\ref{eq.exp.approx}), dashed blue curve: eq.~(\ref{eq.inv2}). The white area indicates the bounds obtained from numerically determined roots of the exact expression for ${G}_\phi(x)$.
(b,d) Numerical solutions of the respective system for different values of $\lambda$ and $\tau$ as indicated.
Shaded areas indicate regions outside the bounds in all plots. The initial condition for all systems is $x|_{-\tau<t<0}=0$.
}
\label{fig.example}
\end{figure*}

\section{Specific delay systems as examples}

We now illustrate this result using specific systems as examples.
The first system is a Mackey--Glass-type system, characterised by a feedback function of the Hill type~\cite{mackeyglass77,novak08},
\begin{equation}
	\phi_n(x) = \frac{2\lambda}{1+x^n} \ , \label{eq.hill}
\end{equation}
where $\lambda$ is the feedback strength and $n$ is the Hill exponent, which determines the nonlinearity of the feedback.
Here, we consider the case $n=2$, the smallest integer value that enables oscillations.
Determining the largest and smallest roots of the function $G_{\phi_2}$ as defined in eq.~(\ref{eq.psi}) yields
\begin{equation}
	x^\pm = \lambda \pm \sqrt{\lambda^2-1} \ , \label{eq.hill.amp}
\end{equation}
which requires $\lambda >1$ for $x^\pm$ to be real.
The invariant point $x^*=\phi_2(x^*)$ between the two amplitude bounds is given by
\begin{equation}
	x^* = \beta - (3\beta)^{-1} \ , \label{eq.hill.eq}
\end{equation}
where $\beta = ( \lambda + \sqrt{\lambda^2 + 1/27} )^{1/3}$, see fig.~\ref{fig.example}a.
Numerical solutions to eq.~(\ref{eq.dyn}) with the feedback function (\ref{eq.hill}) illustrate the effectiveness of the amplitude bounds, see figs.~\ref{fig.example}a,b. 
We find consistently that oscillations saturate the bounds for large values of $\tau$.
Note that the bounds only hold for the limit cycle solution but not necessarily for the initial transient, which depends on the chosen initial history ($x|_{-\tau<t<0}=0$ in all examples).

As a second example, we consider an exponential feedback,
\begin{equation}
	\phi(x) = \lambda \mathrm{e}^{-x} \ . \label{eq.exp}
\end{equation}
While this case does not admit an analytical solution for the bounds, a self-consistent approximation for $x^+$ and $x^-$ can be derived for $\lambda \gtrsim 4$ (as shown below) by approximating the function~$G_{\phi}(x)=\smash{\lambda\mathrm{e}^{-\lambda\mathrm{e}^{-x}}}-x$ for small and large values. Expanding the inner and outer exponential, respectively, to first order, we obtain
\begin{equation}
	{G}_{\phi}(x)
	\approx \begin{cases}
		\lambda\mathrm{e}^{-\lambda(1-x)}-x  \\
		\lambda(1-\lambda\mathrm{e}^{-x})-x 
	\end{cases} \ .
	\label{eq.G.approx}
\end{equation}
By solving the respective approximation for ${G}_{\phi}(x)=0$, we obtain the bound estimates
\begin{equation}
	x^- \approx -\frac{W_0(-\lambda^2\mathrm{e}^{-\lambda})}{\lambda} \ , \qquad x^+ \approx \lambda(1-x^-) \ , \label{eq.exp.approx}
\end{equation}
where $W_n$ is the $n$-th branch of the Lambert W function, defined through the relation $W_n(z) \mathrm{e}^{W_n(z)}=z$ for $z\in \mathds{C}$~\cite{corless96}.
Eqs.~(\ref{eq.exp.approx}) require $\lambda > \lambda_0 = -2 W_{-1}(-\mathrm{e}^{-1/2}/2) \approx 3.513$ to yield real results. The invariant point $x^*$ is given by
\begin{equation}
	x^* = W_0(\lambda) \ . \label{eq.inv2}
\end{equation}
 Fig.~\ref{fig.example}c shows the bounds $x^+$ and $x^-$ from numerical solutions for the roots of the exact expression for ${G}_{\phi}(x)$ (white areas) along with the approximations eqs.~(\ref{eq.exp.approx}) for $\lambda>\lambda_0$ (solid curves).
As shown, eqs.~(\ref{eq.exp.approx}) provide an excellent approximation for the bounds for $\lambda \gtrsim 4$.
Again, we illustrate the effectiveness of the bounds in comparison with numerical solutions of eqs.~(\ref{eq.dyn}), see figs.~\ref{fig.example}c,d. While the exponential feedback yields different waveforms than the Hill feedback, the qualitative features of the two systems are similar.

\section{Generalisations}
We have chosen the delay system eq.~(\ref{eq.dyn}) as a paradigmatic example because its limit cycle can be characterised by the trajectory of a single variable $x$ and its feedback nonlinearity is specified by a single function $\phi$.
Whether the same strategy can be applied to systems involving more components and more complex feedbacks depends on
(i) whether their periodic solution can be characterised by a set of common waveforms for all elements and
(ii) whether a single effective feedback function~$\Phi$ can be constructed which summarises the (possibly indirect) feedback of an oscillating element onto itself.
Typically, to show (ii), a slight modification of Proposition~II has to be invoked.
We now demonstrate that, with little effort, this is possible for many oscillator systems that generate an effective delay through coupling of identical elements.

\section{Systems with effective delays}
\begin{figure*}
\onefigure[width=18cm]{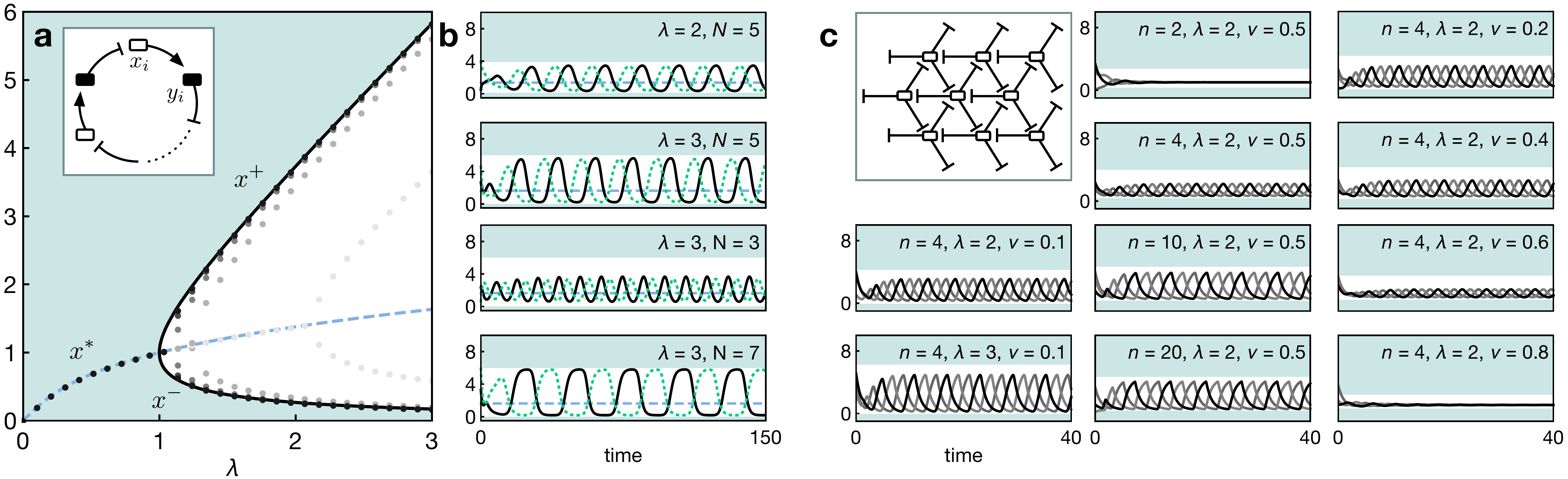}
\caption{(a,b) Analytical bounds and numerical solutions for the repressilator, specified by eqs.~(\ref{eq.intermediates}) with $\psi(x)=x$ and eq.~(\ref{eq.hill}) with $n=2$. All conventions are the same as in Fig.~\ref{fig.example}. (a)~Bounds $x^+$, $x^-$ and invariant point $x^*$ as given by eqs.~(\ref{eq.hill.amp},\ref{eq.hill.eq}); dot colors indicate different $N=3,5,7,11$ from bright to dark. (b) Numerical solutions for different $\lambda$ and $N$ as indicated. For visual clarity, only the variables $x_2$ (solid) and $y_1$ (dotted) are shown.
(c) Analytical bounds and numerical solutions for the repressor lattice system, eqs.~(\ref{eq.rep.lattice},\ref{eq.rep.lattice.cpl}), on a $3\times 3$~lattice with periodic boundary conditions for different parameter sets. For all systems, initial conditions for all variables are randomly drawn from the interval $[0,2\lambda]$.
}
\label{fig.example2}
\end{figure*}

In many biochemical oscillator models, the feedback delay effectively arises through a series of intermediate products~\cite{elowitz00,lewis03,novak08}.
As an example, we consider a ring of repressors,
\begin{equation}
	\dot x_i = \phi(y_{i-1}) - x_i \ , \qquad \dot y_i = \psi(x_i) - y_i  \ , \label{eq.intermediates}
\end{equation}
with $i=1,\hdots,N$.
Here, $y_i$ is the product of $x_i$, which represses $x_{i+1}$, and $\phi$ and $\psi$ are negative and positive feedback functions, respectively ($\phi'<0$, $\psi'>0$), see inset in fig.~\ref{fig.example2}a. We here imply the convention $y_0=y_N$.
The number $N$ of elements determines the effective time delay of the feedback of $x_i$ onto itself. Note that $N$ needs to be odd to ensure a net negative feedback.

For any periodic solution for which all $x_i$ and $y_i$ display the same, possibly time-shifted waveform, i.e., $x_i(t)=\tilde x(t-\delta_i)$ and $y_i(t)=\tilde y(t-\bar\delta_i)$ for appropriate shifts $\delta_i$ and $\bar\delta_i$, we define $\mathcal{C}_\mathrm{X}= \{ \tilde x(t) \ | \ t \}$ and $\mathcal{C}_\mathrm{Y}= \{ \tilde y(t) \ | \ t \}$.
In this case, a straightforward modification of Proposition~II of the proof applied to eqs.~(\ref{eq.intermediates}) yields $\mathcal{C}_\mathrm{X} \subseteq \phi(\mathcal{C}_\mathrm{Y})$ and $\mathcal{C}_\mathrm{Y} \subseteq \psi(\mathcal{C}_\mathrm{X})$. From $\phi'<0$, it then follows that $\phi(\mathcal{C}_\mathrm{Y}) \subseteq \Phi(\mathcal{C}_\mathrm{X})$, where
\begin{equation}
	\Phi(x) = \phi(\psi(x)) \ ,
\end{equation}
and hence $\mathcal{C}_\mathrm{X} \subseteq \Phi(\mathcal{C}_\mathrm{X})$. Now Proposition~III can be applied, i.e., bounds on $\mathcal{C}_\mathrm{X}$ are determined by the effective feedback function~$\Phi$ if it satisfies the appropriate requirements.

To illustrate this result, we choose a repressilator system~\cite{elowitz00} with $\phi$ given by the Hill function $\phi_2$, defined by eq.~(\ref{eq.hill}), and $\psi(x)=x$. Since in this case, $\Phi(x)=\phi_2(x)$, we obtain the same amplitude bounds eq.~(\ref{eq.hill.amp}) as for the delay system eq.~(\ref{eq.dyn}); note however that both systems are different both with respect to their dimensionality and their feedback structure. Since $\mathcal{C}_\mathrm{Y} \subseteq \psi(\mathcal{C}_\mathrm{X})=\mathcal{C}_\mathrm{X}$, the $y_i$ are subject to the same bounds as the $x_i$. Numerical examples confirm the effectiveness of the bounds as shown in figs.~\ref{fig.example2}a,b.
We find that oscillations saturate as $N$ becomes larger, i.e., for large effective delay times.

\section{Spatially extended systems}
Another class of systems that effectively generate delays through coupling are spatially extended systems of the type~\cite{jensen09,chakraborty12}
\begin{equation}
	\dot x_i = \gamma(x_{\langle i1\rangle},\hdots,x_{\langle ik\rangle}) - x_i \ . \label{eq.rep.lattice}
\end{equation}
with $i=1,\hdots,N$ elements, where the $\langle ij\rangle$ are the indices of the $j=1,\hdots,k$ elements that couple into element~$i$. Here, $\gamma$ is the coupling function which we again require to be repressive, i.e., monotonically decreasing in all arguments.
As before, we consider periodic solutions for which all elements are described by a common waveform $x_i(t)=\tilde x(t-\delta_i)$ with appropriate time shifts $\delta_i$ and define $\mathcal{C}= \{ \tilde x(t) \ | \ t \}$. From a slight modification of Proposition~II applied to eq.~(\ref{eq.rep.lattice}), we obtain $\mathcal{C} \subseteq \mathcal{D}_i$ with $\mathcal{D}_i= \{ \gamma(\tilde{x}(t-\delta_{\langle i1\rangle}),\hdots,\tilde{x}(t-\delta_{\langle ik\rangle})) \ | \ t \}$ and from the monotonicity properties of $\gamma$, it follows that $\mathcal{D}_i \subseteq \Phi(\mathcal{C})$ where
\begin{equation}
	\Phi(x)=\gamma(x,\hdots,x) \ ,
\end{equation}
so that $\mathcal{C} \subseteq \Phi(\mathcal{C})$. The fact that $\Phi$ is constructed by evaluating $\gamma$ at equal arguments implies that the derived bounds are based on an in-phase synchronised scenario ($\delta_i = 0$). This may yield quite conservative bounds for dynamically allowed states with $\delta_i \neq 0$. Again, since we have shown $\mathcal{C} \subseteq \Phi(\mathcal{C})$, Proposition~III can be applied if $\Phi$ satisfies the appropriate requirements.

As an example, we consider a hexagonal repressor lattice with multiplicative coupling and periodic boundary conditions, as presented in Refs.~\cite{jensen09,chakraborty12,mengelpers12}, see top left plot in fig.~\ref{fig.example2}c. Coupling is specified by the function
\begin{equation}
	\gamma(x_1,x_2,x_3) = \nu + \frac{\phi_n(x_1)\phi_n(x_2)\phi_n(x_3)}{4\lambda^2} \ , \label{eq.rep.lattice.cpl}
\end{equation}
where $\nu$ is a constant gain rate and $\phi_n$ is the Hill function eq.~(\ref{eq.hill}). Clearly, $\partial \gamma/\partial x_j <0$ for $j=1,2,3$ as required and we obtain $\Phi(x)=\nu + (\phi_n(x))^3/(4\lambda^2)$ which satisfies $\Phi' \leq 0$. Since the resulting roots of $G_\Phi$ do not have closed expressions, we approximate the relevant roots $x^-$ and $x^+$ by considering the first order expansion of $G_\Phi$ at $x=0$  and its asymptotic behaviour for $x\to\infty$, respectively. Setting the respective approximation to zero, we obtain the closed expressions
\begin{alignat}{1}
	x^- &\approx \nu + 2\lambda/(1+(2\lambda+\nu)^n )^{3} \ , \\
	x^+ &\approx \nu + 2\lambda/(1+\nu^n)^{3} \ .
\end{alignat}
As demonstrated in fig.~\ref{fig.example2}c, the obtained approximations are viable bounds for all periodic solution types that fulfil the requirements stated in the previous paragraph. (Here we disregard other possible solution types~\cite{jensen09,chakraborty12}.) 

\section{Discussion}
Our results show how the nonlinear properties of oscillating systems   constrain the amplitude of periodic solutions.
The parameter dependence of the derived amplitude bounds is exclusively governed by the details of the nonlinear feedback---in particular, they are independent of explicit or effective feedback time delays.
Therefore, they hold close to and far away from Hopf bifurcations encountered by varying the time delay.
Our results are valid under very general circumstances, only requiring  certain regularity and monotonicity conditions for the functions describing the negative feedback.
For all considered example systems, we obtained exact bounds or viable analytical approximations and demonstrated their effectiveness in comparison with numerical simulations.
These also showed that the oscillation amplitudes saturate the bounds as the feedback delay becomes much larger than the decay time of the components.
Our results provide a practical method that can be generalised to other biochemical oscillator systems, providing insights into their functional principle and supporting modelling efforts.

\acknowledgments
I thank
\textsc{J.~F.~J\"org} for insightful remarks,
\textsc{S.~Ares}, \textsc{L.~G.~Morelli}, and \textsc{I.~M.~Lengyel} for comments,
\textsc{B.~D.~Simons} for general support,
and an anonymous Referee for valuable suggestions.
I acknowledge the support of the Wellcome Trust (grant number 098357/Z/12/Z).

\pagebreak 

\end{document}